\documentclass[a4paper,12pt]{article}

\usepackage{graphicx}

\usepackage[numbers]{natbib}

\usepackage[latin1]{inputenc}
\usepackage{amssymb,latexsym,amsmath,amscd,epsfig,amsfonts}
\usepackage{epstopdf}
\usepackage{url}
\usepackage{enumerate}

\usepackage{bm}
\usepackage{xcolor}
\usepackage{subfig}
\usepackage{graphicx}
\usepackage[space]{grffile}

\usepackage{algorithm}
\usepackage{algorithmic}

\usepackage{calc}

\usepackage{url}

\usepackage{authblk}
\usepackage{mathrsfs}
\usepackage{tikz}

\newcommand{\h}{$\mathcal H$}
\newcommand{\R}{{\mathbb R}}

\newcommand{\x}{{\bf x}}

\begin{document}
\label{firstpage}
\author[1]{Kjersti Solberg Eikrem}
\author[1,2]{Geir N\ae vdal}
\author[1,3]{Morten Jakobsen}
\affil[1]{NORCE Norwegian Research Centre AS, Postboks 22 Nygårdstangen, 5838 Bergen, Norway.}
\affil[2]{Department of Energy and Petroleum Engineering, University of Stavanger, Norway.}
\affil[3]{Department of Earth Science, University of Bergen, Postboks 7803, 5020 Bergen, Norway.}
\date{}
\renewcommand\Affilfont{\itshape\small}

\date{}

\title{Iterative solution of the Lippmann-Schwinger equation in strongly scattering acoustic media by randomized construction of preconditioners}
\maketitle 

\begin{abstract}
In this work  the Lippmann-Schwinger equation is used to model seismic waves in strongly scattering acoustic media. We consider the Helmholtz equation, which is the scalar wave equation in the frequency domain with constant density and variable velocity, and transform it to an integral equation of the Lippmann-Schwinger type. To directly solve the discretized problem with matrix inversion is time-consuming, therefore we use iterative methods. The Born series is a well-known scattering series which gives the solution with relatively small cost, but it has limited use as it only converges for small scattering potentials. There exist other scattering series with preconditioners that have been shown to converge for any contrast,  but the methods might require many iterations for models with high contrast. 
Here we develop new preconditioners based on randomized matrix approximations and hierarchical matrices which can make the scattering series converge for any contrast with a low number of iterations. We describe two different preconditioners; one is best for lower frequencies and the other for higher frequencies. We use the fast Fourier transform (FFT) both in the construction of the preconditioners and in the iterative solution, and this makes the methods efficient. The performance of the methods are illustrated by numerical experiments  on two 2D models.
\end{abstract}




\section{Introduction}
The Lippmann-Schwinger equation can be used to describe many physical phenomena, for example acoustic and electromagnetic scattering of waves and scattering of particles in quantum physics \cite[]{lippmann1950variational,taylor1972scattering,clayton1981born}. In this paper we use the Lippmann-Schwinger equation to 
model seismic waves in strongly scattering media. We consider the Helmholtz equation, which is the scalar wave equation in the frequency domain, and transform it to an integral equation of the Lippmann-Schwinger type. 
To directly solve the linear system resulting from the discretization of the problem is time-consuming, and therefore iterative solutions are more advantageous. A simple iterative solution is the Born series \cite[]{morse1953methods}, which converges only for small contrasts. Recently, other scattering series with better convergence properties have been studied \cite[]{osnabrugge2016convergent,eftekhar2018convergence,huang2019taming,jakobsen2020homotopy}. In 
\cite{osnabrugge2016convergent} a scattering series with a preconditioner was used to solve the Helmholtz equation for light propagation. The same method was tested for seismic modelling in \cite{huang2019taming}, and a generalization of this series based on the homotopy analysis method \cite[]{liao2003beyond} was obtained in \cite{jakobsen2020homotopy}. The series in \cite{osnabrugge2016convergent} was proven to converge for a particular choice of preconditioner, but convergence could be slow for large scattering potentials, as is often the case in seismic applications. In general the convergence speed of these series depends on the quality of the preconditioner, and in this work we develop methods for obtaining preconditioners by the use of randomized methods and hierarchical matrices.

Randomization is a powerful tool for performing large-scale matrix operations more efficiently. In many cases,  randomized algorithms can be faster and more stable than classical algorithms \cite[]{halko2011finding,drineas2016randnla}.
Recently, the usefulness of randomized methods has been demonstrated on many applications. In \cite{isaac2015scalable} randomized singular value decomposition (SVD) was used in algorithms for inversion and prediction of flow of the Antarctic ice sheet. Randomized data reduction was used in \cite{lin2017large} to invert for the transmissivity field in groundwater flow. 
In \cite{bjarkason2018randomized} a Levenberg-Marquardt method with randomized truncated singular value decomposition was used for history matching of a geothermal reservoir.

Hierarchical matrices are approximations of full matrices. The approximations are done block-wise, by dividing the matrix according to a tree structure, and using low rank approximations for many of the blocks. Hierarchical matrices were introduced in \cite{hackbusch1999sparse}, and have since found many applications. In particular, such matrices can be used as preconditioners to solve many different equations. In for example \cite{banjai2008hierarchical,chaillat2017theory} hierarchical matrices were used as preconditioners to solve the Helmholtz equation with the boundary element method, and in \cite{chaillat2017theory} also  the elastodynamic equation was solved. In \cite{engquist2011sweeping}  the Helmholtz equation was solved with  the finite difference method and a hierarchical preconditioner. 

In this work we demonstrate two ways of obtaining preconditioners for the scattering series. The first method is only based on randomized approximations  of the matrix we need to invert, and works well for smaller examples and lower frequencies. In the second method the approximations are done in a hierarchical way, but still using randomized methods. This approach works better for the larger models and higher frequencies. We use the fast Fourier transform (FFT) both in the construction of the preconditioners and in the iterative solution to speed up matrix-vector multiplication.

In this paper we have focused on using the preconditioners with convergent scattering series, but the same preconditioners can also be applied to Krylov subspace methods. We compare the performance of the scattering  series with GMRES with and without a preconditioner.  
Our interest in scattering series is not only because of computational speed in the forward modelling, but also because it could lead to further developments in inverse scattering series which could be usedful for inversion, see \cite{weglein2003inverse,kouri2003inverse}. Convergence of the forward scattering series does not necessarily imply that the inverse scattering series converges, but this is outside the scope of this paper.

The Helmholtz equation,  which we consider in this work, is the scalar wave equation in the frequency domain for acoustic media with variable velocity but constant density. The scalar wave equation can be regarded as an approximation to the acoustic wave equation with variable density and compressibility,
 which is turn can be regarded as an approximation to the (anisotropic) elastodynamic wave equation. The scalar wave equation is sometimes used in exploration seismology in the context of full waveform inversion and seismic imaging because it reduces the computational cost compared to similar approaches based on the acoustic wave equation with variable density and the elastodynamic wave equation. We believe the scattering series with preconditioners presented here can be further developed to also work for more general wave propagation and scattering problems. We apply our methods to 2D examples, but with some small changes it will also work for 3D models. 

The outline of the paper is as follows. In section \ref{methods} we describe the methods. The Lippmann-Schwinger equation is shown in section \ref{Helmholtz}, and how it can be solved by scattering series is described in section \ref{scattering section}. The randomized preconditioners are presented in section \ref{random section} and \ref{hierarchical section}. 
The numerical examples are presented in section \ref{numerical}, and the conclusion follows in section \ref{conclusion}.

\section{Theory}
\label{methods}
\subsection{The Lippmann-Schwinger equation}
\label{Helmholtz}
We  assume that the seismic 
wavefield $\psi ({\bf x},\omega)$ at point ${\bf x}$ 
due to a source density $S({\bf x},\omega)$ in a medium with variable velocity
$c({\bf x})$ and constant density
satisfies the Helmholtz equation 
(see \cite{morse1953methods}):                    
\begin{equation*}
\left[ \nabla ^{2} + \frac{\omega ^{2}}{c^{2}({\bf x})}\right] \psi ({\bf x},\omega)  = -S({\bf x},\omega).
\label{wave_equation}
\end{equation*}
Here $\omega$ is the angular frequency. The wavefield $\psi ({\bf x},\omega)$ is given by the following volume 
integral \cite[]{morse1953methods}:
\begin{equation*}
\psi ({\bf x},\omega) = \int G({\bf x}, {\bf x}',\omega)S({\bf x}',\omega)d{\bf x}', 
\label{source_representation}
\end{equation*}
where the integration is over all of the space ( $\R^2$ or  $\R^3$) 
and
$G({\bf x}, {\bf x}',\omega)$ is the Green's function, which is defined by 
\begin{equation*}
\label{gr}
\left[ \nabla ^{2} + \frac{\omega ^{2}}{c^{2}({\bf x})}\right] G({\bf x}, {\bf x}',\omega) = -\delta ({\bf x} - {\bf x}'),
\end{equation*}
where $\delta$ is Dirac's delta function. 
We introduce the contrast $\chi$ relative to an arbitrary homogeneous background medium $c_0$
\begin{equation}
\label{slown}
 \frac{\omega^2}{c^{2}({\bf x})} = \frac{\omega^2}{c^{2}_{0}} + \chi ({\bf x}).
\end{equation} 
Then
 \begin{equation}
\label{contrast source}
\left[ \nabla ^{2} + \frac{\omega ^{2}}{c^{2}_{0}}\right] \psi ({\bf x},\omega)  = -S({\bf x},\omega) - \chi ({\bf x})\psi ({\bf x},\omega).
\end{equation}
The last term on the right-hand side of \eqref{contrast source} represents a contrast-source term which can be treated just like the ordinary source term $S$. As a result, the partial differential equation \eqref{contrast source} can be transformed into an equivalent integral equation of the Lippmann-Schwinger type \cite[]{lippmann1950variational,taylor1972scattering}, 
\begin{equation}
\label{split}
\psi ({\bf x},\omega) = \psi ^{(0)}({\bf x},\omega) + \int_D G^{(0)}({\bf x}' -{\bf x},\omega)\chi ({\bf x}')\psi ({\bf x}',\omega)d{\bf x}',
\end{equation}
where $G^{(0)}$ is the Green's function for the  
background medium, $\psi ^{(0)}$ is the wavefield in the background medium, ${\bf x}$ is any point in space and $D$ is the domain where $\chi $ is nonzero. 
Since the background is homogeneous ($c_0$ is a constant) 
the Green's function for the background medium is translation invariant, 
i.e $G^{(0)}({\bf x}, {\bf x}',\omega)=G^{(0)}({\bf x}' -{\bf x},\omega)$. Therefore \eqref{split} is a convolution integral, and we will make use of that later when we will apply the fast Fourier transform (FFT) to speed up calculations.

We discretize \eqref{split} by using the values in the centres of the grid blocks on a cartesian grid, and get the following matrix equation 
\begin{equation}
	\label{discr}
	\psi = \psi^0+G^0V\psi,
\end{equation}
where $V$ is a diagonal matrix with $ \Delta v \chi$  
on the diagonal, $\Delta v$  
is the volume of a grid block,  
and $G^0$ is $G^{(0)}$ evaluated in the gridblocks. 
We use the same notation $\psi$ for the discretized function as for the continuous function for simplicity.
Equation \eqref{discr} can be rewritten as
\begin{equation}
	\label{solve}
	(I-G^0V)\psi=\psi^0,
\end{equation}
which can  be solved by for example matrix inversion:
\begin{equation}
\label{full}
	\psi = (I -G^0V)^{-1}\psi^0.
\end{equation}
But to calculate the required inverse is time-consuming for large models, and therefore we investigate iterative solutions.

\subsection{Solution by scattering series}
\label{scattering section}
If the contrasts are small, the solution of \eqref{full} can be found using the Born series (see for example \cite{morse1953methods})
\begin{equation*}
	\psi =(I+G^0V+G^0VG^0V+...)\psi^0=\sum_{n =0}^\infty (G^0V)^n \psi^0. 
\end{equation*}
This can be seen by expanding \eqref{discr} recursively. Let
\begin{equation*}
	\psi_j=\sum_{n =0}^j (G^0V)^n \psi^0,
\end{equation*}
then the solution can be found iteratively
\begin{equation*}
	\psi_j =G^{0}V\psi_{j -1}+\psi^0.
\end{equation*}
The series  will converge to $\psi$ if the spectral radius 
(the maximum of the absolute values of the eigenvalues) of $G^{0}V$ is less than 1. For large models this is rarely fulfilled \cite[]{jakobsen2015Full,osnabrugge2016convergent}.

In \cite{osnabrugge2016convergent} a scattering series with a preconditioner was used to solve the Helmholtz equation for light propagation. It was shown	that the series
\begin{equation*}
	\psi =\sum_{n =0}^\infty M^n \gamma\psi^0
\end{equation*}
with $\gamma=iV/\epsilon$ and
$M =I-\gamma+\gamma G^{0}V$
converges as long as  $\epsilon$ is chosen such that 
\begin{equation*}
\epsilon \ge \omega^2 \max_\x \left|\frac{1}{c^{2}({\bf x})}-\frac{1}{c_0^2}\right|.
\end{equation*}
 Here  $G^0$ was modified by introducing dissipation in the background medium, and to remove the effect of $\epsilon$, a gain is added in $V$ , i.e. $V$ has $\chi+i\epsilon$ on the diagonal. Absorbing boundary layers were used to remove artificial reflections because of  $\epsilon$. 
The convergence rate of the series depends on $\epsilon$; the larger $\epsilon$ is, the slower is the convergence. Higher frequencies and stronger contrast in velocity require larger $\epsilon$, and will therefore slow down the convergence rate. 
It is noted in \cite{osnabrugge2016convergent} that the scattering contrast in optical systems is relatively small, and the method was fast for the numerical example in that paper, but in acoustic wave simulations the scattering contrast can become much larger, and therefore reduce the speed of the method.
In \cite{huang2019taming,huang2020applicability} it was demonstrated that the method could also be applied for 
seismic modelling.

 In \cite{jakobsen2020homotopy} similar series were investigated.   
It was shown that if one could find a  matrix (or more generally an operator) 
$H$  such that the spectral radius of 
\begin{equation}
\label{mm}
	M =I-H+HG^{0}V
\end{equation}
was less than 1, then the solution to the Helmholtz equation can be found as a series
\begin{equation}
\label{series hom}
	\psi =\sum_{n =0}^\infty \phi_n,
\end{equation}
where $\phi_0$ is an initial guess of the solution to \eqref{solve}, $\phi_1=H(\psi^0-\phi_0+G^0V\phi_0)$ and $\phi_n=M\phi_{n-1} $ for $n\geq 2$.
H is called a convergence control operator in \cite{jakobsen2020homotopy}, but it can also be viewed as a preconditioner. (In \cite{jakobsen2020homotopy} there was an additional scalar parameter $h$ which was multiplied with $H$, but as we will not work with them independently, we have included it in $H$  except for a sign difference.) The results were obtained through the homotopy analysis method \cite[]{liao2003beyond}, and the series was shown to be a generalization of the series from \cite{osnabrugge2016convergent} in the sense that the series coincide if one  uses $H=\gamma$ and $\phi_0=\gamma\psi^0$. 
The initial guess  $\phi_0$ is  arbitrary, and the series will converge as long as the spectral radius of $M$ is less than one. The preconditioners we construct will work for any initial guess, but in our numerical examples we will use $\phi_0=H\psi^0$. 
Then $\phi_1=H(\psi^0-H\psi^0+G^0VH\psi^0)=MH\psi^0$, and the series can be written
\begin{equation}
\label{series}
	\psi =\sum_{n =0}^\infty M^n H\psi^0. 
\end{equation}
Another way to show that this series will give the solution is to multiply \eqref{discr} with $H$ and add $\psi$ on both sides:
\begin{equation*}
\psi+H\psi=\psi+ H\psi^0+HG^{0}V\psi.
\end{equation*}
Then we can rearrange the formula to obtain
\begin{equation*}
\psi=(I-H+HG^{0}V)\psi+H\psi^0=M\psi+H\psi^0,
\end{equation*}
and by expanding it recursively we obtain \eqref{series}.
If H is chosen such that the spectral radius of M is less than 1, then \eqref{series} will converge. 
Different $H$'s were tested in \cite{jakobsen2020homotopy}, with different speeds of convergence. One choice was $\alpha I$, where $I$ is the identity matrix, and $\alpha$ is a scalar $<1$. Also multiples of the preconditioner $\gamma$ 
from \cite{osnabrugge2016convergent} were tested, and  it was shown that a multiple of $\gamma$ could give faster convergence. 
All the choices for $H$ that were tested were diagonal matrices, and there was no general procedure on how $H$ should be selected. Convergence can be ensured by using $H=\gamma$, 
but the number of iterations could be large.  In this work we find preconditioners 
that can reduce the number of iterations  by also considering non-diagonal $H$. 

Similarly as for the Born series, if we define 
\begin{equation*}
	\psi_j =\sum_{n =0}^j M^n H\psi^0,
\end{equation*}
then the solution can be found iteratively
\begin{equation}
	\psi_j =M\psi_{j -1}+H\psi^0
	\label{update m}
\end{equation}
for $j>1$ and $\psi_0 =H\psi^0$.
The updating formula \eqref{update m} can be rearranged in the following way 
\begin{equation}
\label{update}
	\psi_j =\psi_{j -1}-H(\psi_{j -1}-G^0V\psi_{j -1}-\psi^0)
\end{equation}
by using the definition of $M$ in \eqref{mm}. 
 To calculate this in a fast way, we use FFT (as was also done in \cite{osnabrugge2016convergent}). The product $G^0(V\psi_{j-1})$ can be calculated efficiently using FFT because of the structure of the Green's function. For a homogeneous background the Green's function $G^0$ is a block-Toeplitz matrix, and therefore FFT can be used, see for example \cite{nowak2003efficient}. A good explanation is also given in \cite{mojabi2015ultrasound}. Another way to see that FFT can be used, is that the integral in \eqref{split} is a convolution when the background is homogeneous, and then one can calculate the pointwise multiplication in the Fourier domain and then do the inverse Fourier transform of the result. 
We calculate
\begin{equation*}
\mathscr{F}^{-1}(\mathscr{F}(\tilde{G^0}) \odot \mathscr{F}(V\psi_{j-1}))
\end{equation*}
where $ \odot$ denotes pointwise multiplication and $\mathscr{F}$ is the two-dimensional FFT  when we work in 2D, but a similar procedure can be done in 3D. 
We wrote a $\sim$ over $G^0$ to emphasize that it is not the full matrix $G^0$ that is used, but only the first row of the matrix, reshaped as a matrix of size $N_x \times N_y$ which is the size of the numerical model, and then extended as described in \cite{nowak2003efficient}. Also the diagonal of $V\psi_{j-1}$ is reshaped, and extended with zeros. 


\subsection{Approximations by randomized methods}
\label{random section}
We want the spectral radius of $M$ in \eqref{mm} to be as small as possible for fast convergence. Heuristically, $M$ should be close to 0, and that will happen if $H(I-G^0V) \approx I$, i.e. 
\begin{equation*}
H \approx (I-G^0V)^{-1}. 
\end{equation*}
To obtain a good approximation of  $ (I-G^0V)^{-1}$,  we will use 
randomized algorithms. 
First we will show a method where we compute a low rank approximation of the matrix $G^0V$, and then a method where $ (I-G^0V)^{-1}$ is approximated by a hierarchical matrix. The first method works best for lower frequencies, and the second for higher frequencies, so we will describe both. The simple method also has the advantage of being very easy to implement, and it is a buildingblock in the algorithm with hierarchical matrices. 

If we obtain an approximation of $G^0V$ by a product of two low rank matrices,
\begin{equation*}
	G^0V\approx UW^T,
\end{equation*}
where $U$ and $W$ are of dimensions $n \times r$ with $r <<n$, it is easy to find an approximation of  $ (I-G^0V)^{-1}$. The following matrix identity is the Sherman-Morrison-Woodbury formula 
\begin{equation*}
	(A-BC)^{-1}=A^{-1}+A^{-1}B(I-CA^{-1}B)^{-1}CA^{-1},
\end{equation*}
and it holds if $A$ and $(I-CA^{-1}B)$ are invertible (see for example \cite{Golub86Matrix,hager1989updating}).
By using this identity, we get
\begin{equation*}
	(I_n-G^0V)^{-1}\approx (I_n-UW^T)^{-1}=I_n+U(I_r-W^TU)^{-1}W^T,
\end{equation*}
where the subscript of $I$ indicates the dimension of the identity matrix and $T$ denotes the complex conjugate transpose (as we work with complex matrices). Then we choose
\begin{equation}
\label{hrk}
H=I_n+U(I_r-W^TU)^{-1}W^T
\end{equation}
Note that  
$(I_r-W^TU)$ is of dimension $r \times r$ with $r<<n$, and therefore cheap to invert. The matrix $H$ is no longer a diagonal matrix as in \cite{osnabrugge2016convergent} and \cite{jakobsen2020homotopy}, and to avoid large computational cost when multiplying vectors with $H$, the product $U(I_r-W^TU)^{-1}W^T$ should not be performed, but kept as three separate factors. We only calculate $Z=(I_r-W^TU)^{-1}$. 
The update is then performed in two steps,
\begin{equation}
\label{update2}
	a_{j-1}=\psi_{j -1}-G^0V\psi_{j -1}-\psi^0
\end{equation}
and
\begin{equation}
\label{update3}
	\psi_j =a_{j -1}+U(Z(W^Ta_{j-1})).
\end{equation}
In this way we calculate the product of a vector $a_{j-1}$ times three low rank matrices instead of a vector times a full matrix, as we would have if we calculated $H=I+UZW^T$ in advance. The product $G^0(V\psi_{j-1})$ is calculated using FFT. The accuracy of the method is high as one can reach machine precision in few iterations as long as the spectral radius of $M$ is small enough.

We use an algorithm  from \cite{halko2011finding} to obtain an approximate decomposition of $G^0V$. For clarity we state the original algorithm first, and then show the modifications we use to make it faster for our application. Algorithm 4.4 from \cite{halko2011finding} can be used to find the approximate range of a matrix. Algorithm \ref{halko} shows an extended version of this algorithm. (Note that point 5 is not in the original algorithm, but is mentioned elsewhere in the paper. We added it for completeness.) 

\begin{algorithm}[H]
\begin{algorithmic} 
 \STATE Given an $m\times n$ matrix A and and integers $r>0$  and $q\ge 0$.
\STATE 1. Draw an $n\times r$ Gaussian random matrix $\Omega$ (a matrix of numbers from the standard normal distribution).
\STATE 2. Form the $m\times r $ matrix $Y_0=A \Omega$.
\STATE 3. Construct an $m\times r $ matrix  whose columns form an orthonormal basis for the range of $Y_0$, e.g. using QR factorization $Y_0=Q_0R_0$.
\STATE 4. for j=1:q\\
\quad	Form $\tilde{Y}_j=A^TQ_{j-1}$ and compute the QR factorization $\tilde{Y}_j=\tilde{Q}_j\tilde{R}_j$.\\
\quad	Form $Y_j=A\tilde{Q}_{j}$ and compute the QR factorization $Y_j=Q_jR_j$.\\
end
\STATE 5. Form $\tilde{Y}_{q+1}=A^TQ_{q}$. \\
Then $Q_q\tilde{Y}_{q+1}^T$ is an low rank approximation of A.
\end{algorithmic} 
\caption{Algorithm for constructing a low rank approximation of a matrix $A$, from \protect\cite{halko2011finding}. }
\label{halko}
\end{algorithm}
The simplest version of Algorithm \ref{halko} is to use $q=0$, and then point 4 in the algorithm is skipped. Using $q>0$ can be beneficial for increasing the accuracy, especially for large matrices and for matrices where the singular values decay slowly, but the cost of the algorithm will be larger as well. 

The bottleneck of  Algorithm \ref{halko} is usually to calculate the product of A or $A^T$ with $\Omega$ and $Q$, but in our case we can make use of the structure of $G^0$ to do these calculations much faster using FFT, see Algorithm \ref{random algorithm}.  The approximation we obtain is used in the update formulas \eqref{update2} and \eqref{update3} with $U=Q_q$ and $W=\tilde{Y}_{q+1}$.

\begin{algorithm}
\begin{algorithmic} 
\caption{Algorithm for constructing a low rank approximation of the matrix $G^0V$. }
\label{random algorithm}
 \STATE Given the Green's function $G^0$, the diagonal $n\times n$ matrix $V$ and integers $r>0$ and $q\geq 0$.
\STATE 1. Draw an $n\times r$ Gaussian random matrix $\Omega$  and calculate $V\Omega$.
\STATE 2. Calculate the $n\times r $ matrix $Y_0=G^0(V\Omega)$ using FFT.
\STATE 3. Construct an $n\times r $ matrix  whose columns form an orthonormal basis for the range of $Y_0$, e.g. using QR factorization $Y_0=Q_0R_0$.
\STATE 4. 
for j=1:q \\
\quad		Calculate  ${G^{0T}}Q_{j-1}$ using FFT, let $\tilde{Y_{j}}=V{G^{0T}}Q_{j-1}$. Find the QR factorization $\tilde{Y_j}= \tilde{Q_j}\tilde{R_j}$.\\
\quad		Calculate  $VQ_j$ and then ${Y_j}=G^0V\tilde{Q_j}$ using FFT, and find the QR factorization ${Y_j}={Q_j}{R_j}$. \\
		end
\STATE 5. 
Calculate  ${G^{0T}}Q_{q}$ using FFT and let $\tilde{Y}_{q+1}=V{G^{0T}}Q_{q}$. \\
Then $Q_q\tilde{Y}_{q+1}^T$ is an low rank approximation of $G^0V$.
\end{algorithmic} 
\end{algorithm}

We will see in the numerical examples later that this  method with $H$ as in \eqref{hrk} works well for lower frequencies and small models, but for higher frequencies it is better to use hierarchical matrices, which we will describe in the next section.


\subsection{Randomized construction of hierarchical matrices}
\label{hierarchical section}
Hierarchical matrices (also called $\mathcal{H}$-matrices) are  data-sparse approximations of non-sparse matrices. The matrices are not sparse in the sense that they contain a lot of zeros, but they are divided in blocks based on a tree structure, and most of the blocks are represented by low rank matrices \cite[]{hackbusch1999sparse,borm2003hierarchical}. If $R$ is a sub block  of dimension $m\times n$, it can be approximated by a product of two low rank matrices,
	$R=AB^T$,
 where $A $ has dimension $m\times r$ and B has dimension $n\times r$. Which blocks will be kept as full matrices and which will be approximated is decided in advance based on a priori knowledge of the matrix and how the degrees of freedom are ordered.

 We will approximate $ (I-G^0V)  $ by a \h-matrix and then find an approximate inverse, which is also a  hierarchical matrix. This inverse will be used as H in \eqref{update}.

\begin{figure}
\centering
\begin{tikzpicture}

\foreach \x in {0,0.125,...,3.875} 
\filldraw[color=black,fill=black!20] (0+\x,3.875-\x) rectangle (0.125+\x,4-\x);
\draw (0,0) rectangle (4,4);

\foreach \x in {0,2}
\draw (0+\x,0+\x) rectangle (2+\x,2+\x);

\foreach \x in {0,1}
\draw (0+\x,2+\x) rectangle (1+\x,3+\x);
\foreach \x in {0,1}
\draw (2+\x,0+\x) rectangle (3+\x,1+\x);

\foreach \x in {0,0.5}
\draw (0+\x,3+\x) rectangle (0.5+\x,3.5+\x);
\foreach \x in {0,0.5}
\draw (1+\x,2+\x) rectangle (1.5+\x,2.5+\x);
\foreach \x in {0,0.5}
\draw (2+\x,1+\x) rectangle (2.5+\x,1.5+\x);
\foreach \x in {0,0.5}
\draw (3+\x,0+\x) rectangle (3.5+\x,0.5+\x);

\foreach \x in {0,0.25}
\draw (0+\x,3.5+\x) rectangle (0.25+\x,3.75+\x);
\foreach \x in {0,0.25}
\draw (1+\x,2.5+\x) rectangle (1.25+\x,2.75+\x);
\foreach \x in {0,0.25}
\draw (2+\x,1.5+\x) rectangle (2.25+\x,1.75+\x);
\foreach \x in {0,0.25}
\draw (3+\x,0.5+\x) rectangle (3.25+\x,0.75+\x);

\foreach \x in {0,0.25}
\draw (0.5+\x,3+\x) rectangle (0.75+\x,3.25+\x);
\foreach \x in {0,0.25}
\draw (1.5+\x,2+\x) rectangle (1.75+\x,2.25+\x);
\foreach \x in {0,0.25}
\draw (2.5+\x,1+\x) rectangle (2.75+\x,1.25+\x);
\foreach \x in {0,0.25}
\draw (3.5+\x,0+\x) rectangle (3.75+\x,0.25+\x);
\end{tikzpicture}
\caption{The figure illustrates the  
decomposition into submatrices that is used for finding a hierarchical matrix to approximate $ I-G^0V  $. The off-diagonal blocks (white) are approximated by  products of two low rank matrices, and the diagonal blocks (grey) are kept as dense matrices.}
\label{matrix}
\end{figure}
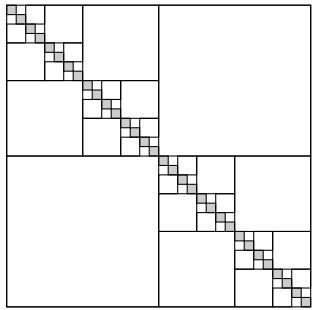

Our grid is ordered 
 columnwise from left to right, i.e in vertical strips.
We use a simple structure for the hierarchical matrix, by approximating all off-diagonal blocks. This particular structure of the hierarchical matrix is denoted hierarchically off-diagonal low-rank (HODLR). 
When finding an approximation for $ (I-G^0V)  $, we start by dividing the matrix in four. Then the two off-diagonal blocks are approximated  as in Algorithm \ref{random algorithm}, and the blocks on the diagonal are further divided in four, and the procedure is repeated, see Fig. \ref{matrix}.
The off-diagonal blocks of $ G^0V  $ corresponds to scattering between vertical slices of the model. After the first division in four, we approximate the scattering between the left and right half of the model by using low-rank matrices for the two off-diagonal blocks, and the next division approximates the scattering between the left and right quarters after each half is divided in two, and so on.

The same rank  $r$ is used for all subblocks. 
 The matrix $G^0$ has blocks of size $N_y\times N_y$ where $N_y$ is the number of grid blocks in the vertical direction, and when we divide, we construct blocks that are multiples of $N_y\times N_y$. Then the blocks are not necessarily equal in size, but  the pattern is followed, and that makes it easier to use FFT. 
(The grid could have been organized horizontally such that $G^0$ had blocks of size $N_x\times N_x$ instead, but as the models we work which are longer in the horizontal direction, we get larger blocks and slower decay of the Green's function outside the diagonal, which resulted in larger computational time due to the need for a higher rank $r$.)
We continue until we reach a minimum size of the blocks (or maximum number of levels) which is chosen in advance. The remaining blocks on the diagonal are kept as full matrices, and not approximated.
The matrix $ (I-G^0V)  $  has $N^2$ elements, where $N$ is the number of grid blocks in the model, hence for large models it is very costly to store. Therefore we do the approximations without explicitly forming the matrix $ (I-G^0V)  $. We only form the sub-blocks that are used on the diagonal explicitly. The off-diagonal blocks are approximated using FFT to speed up the calculations.

The formula 
\begin{equation*}
	\begin{pmatrix}
	A & B\\
	C &D 
	\end{pmatrix}^{-1} =
		\begin{pmatrix}
	A^{-1}+A^{-1}B(D-CA^{-1}B)^{-1}CA^{-1} & -BA^{-1}(D-CA^{-1}B)^{-1}\\
	-(D-CA^{-1}B)^{-1}CA^{-1} &(D-CA^{-1}B)^{-1} 
	\end{pmatrix}
\end{equation*}
shows how a block matrix can be inverted  \cite[]{hackbusch1999sparse}. We use this formula recursively when performing the inversion of the hierarchical matrix. The inverted matrix has the same structure as shown in Fig. \ref{matrix}.

When performing the inversion we need to perform addition and multiplication of the sub-matrices. 
When two blocks are added several cases can occur. If two blocks of full matrices are added, the addition is the usual addition of matrices. If two low-rank approximations of rank $r$ are added, one either has to increase the rank to $2r$ or do an approximation to keep the rank as $r$ \cite[]{hackbusch1999sparse,boerm03introduction}. Singular value decomposition can be used to find the best approximation of rank $r$. Here we use randomization also in the addition of matrices to speed it up. It is not as accurate as the deterministic singular value decomposition, but faster. We use a slightly modified version of Algorithm \ref{halko} blockwise to reduce the rank after addition. We make use of the fact that the block is of low rank to do the multiplication faster, i.e. if a block is $A=BC^T$, we calculate $B(C^T\Omega)$ instead of $A\Omega$  in Algorithm \ref{halko}, and similarly for the $Q$'s. We used $q=0$ in Algorithm \ref{halko} and \ref{random algorithm} for the hierarchical decomposition and inversion.



\section{Numerical experiments}
\label{numerical}
We first test the methods on a relatively small model to show that it gives the same result as solving equation \eqref{full} directly. Afterwards we show that the methods can also be applied on a larger example, where using equation \eqref{full} would be very time-consuming and memory demanding. The code is implemented in MATLAB, and we used a desktop computer with CPU speed of 3.4 GHz.


\subsection{Verification of the methods}
\begin{figure*}
\centering
\includegraphics[width=0.6\textwidth]{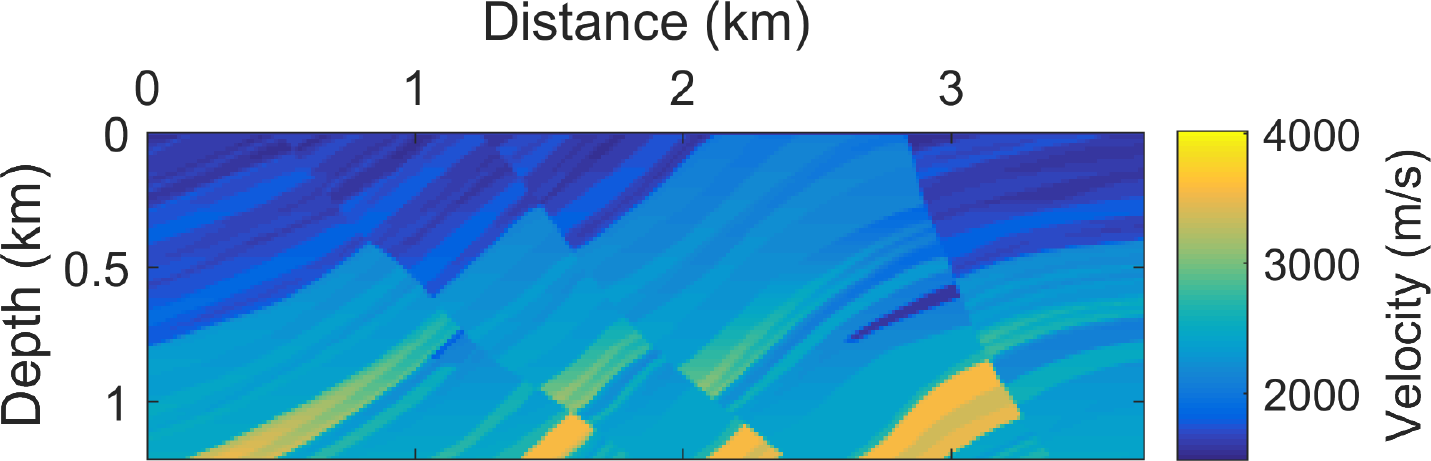}
\caption{A subset of the Marmousi2 model.}
\label{small model}
\end{figure*}

As the first test model we use a resampled subset of the Marmousi2 model \cite[]{martin2006marmousi2}, see Fig. \ref{small model}. The model has $248\times 81= 20088$ grid blocks of size 15 m in both directions. We assume the surroundings of the model have velocity 2000 m/s, and this is used as $c_0$ in \eqref{slown}. For this model the Born series only converges for 1 and 2 Hz. For higher frequencies the contrast is too large and we need preconditioners to make the scattering series convergent.
We will test both the two preconditioners described above.

We use the integer frequencies from 1 - 20 Hz. The source is a Ricker wavelet with centre frequency 10 Hz, and it is placed in the middle at the top of the model.
When performing the forward simulation for several frequencies, we start with the lowest, as it is the easiest to approximate. For higher frequencies we need a larger rank for the approximations because of more oscillations in the Green's functions. 

To calculate the spectral radius of $M$ is time-consuming for large matrices, so we do not do that. Instead we just test whether we have convergence of \eqref{update} within a fixed number of iterations. We used 30 as the upper limit. If we do not have convergence within this number, we recalculate the preconditioner  
with a larger number for $r$ and restart the iterations from the original $\psi^0$. If convergence was obtained, but  more than 10 iterations were needed, 
we increase $r$ for the next frequency. In this way we mostly avoid recalculations. How much $r$ needs to be increased is case dependent, but a few experiments will give a suitable value. We used $||\psi_j-\psi_{j-1}||<10^{-5}$ as stopping criteria when updating with formula  \eqref{update2} and \eqref{update3} for the simple preconditioner and \eqref{update} for the hierarchical preconditioner.

\subsubsection{Simple preconditioner}
\begin{figure*}
\centering
\subfloat[Real part of the solution obtained by solving~\eqref{full}.]{
\includegraphics[width=0.48\textwidth]{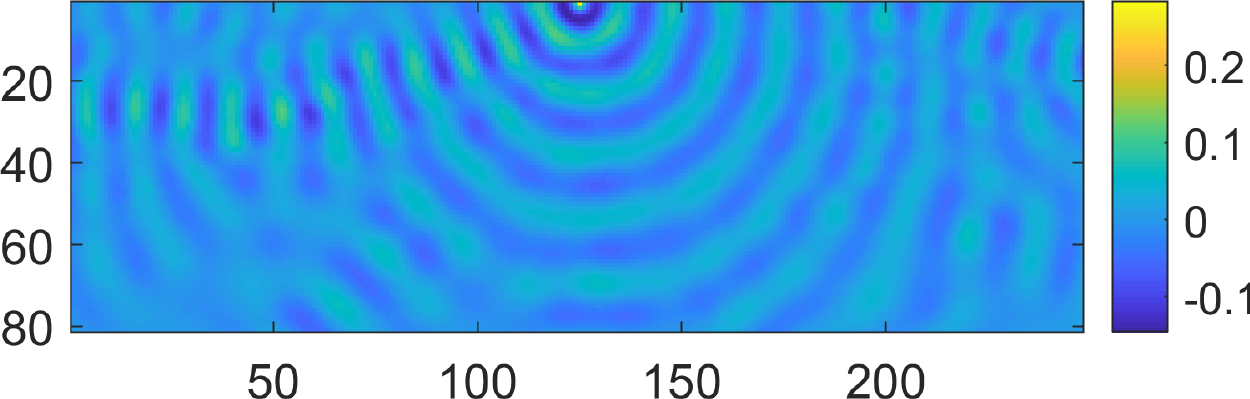}
\label{estimated difference 1}
}
\subfloat[Imaginary part of the solution obtained by solving~\eqref{full}.]{
\includegraphics[width=0.48\textwidth]{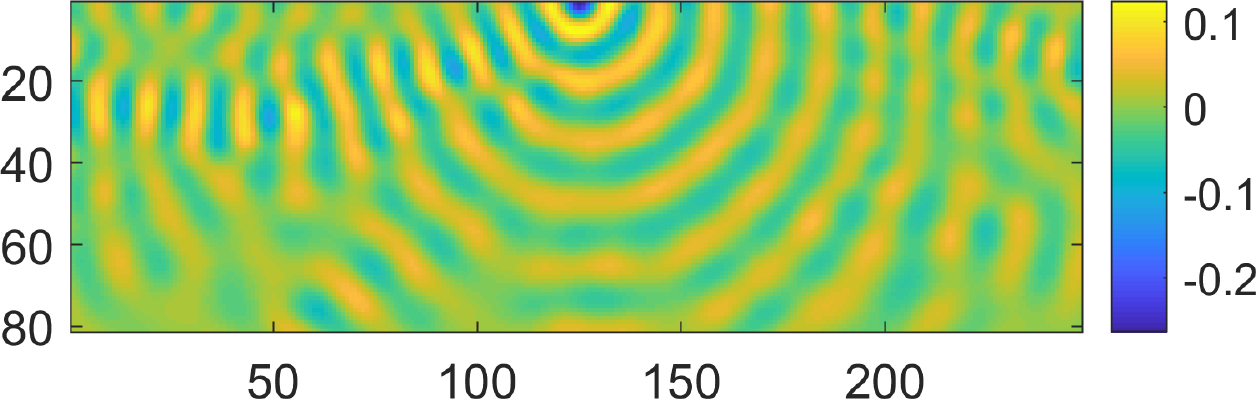}
\label{error diff 1}
}\\
\subfloat[Real part of the solution obtained by using the simple randomized preconditioner.]{
\includegraphics[width=0.48\textwidth]{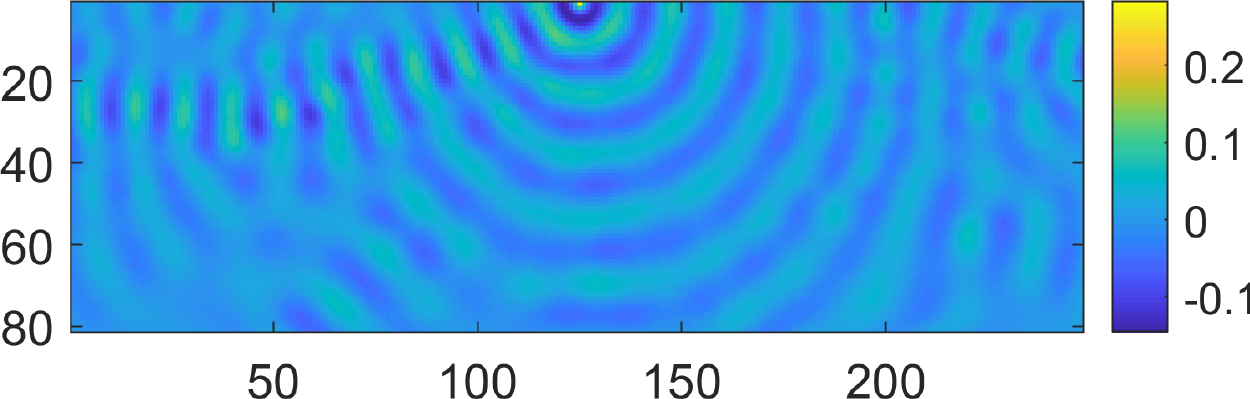}
\label{estimated difference 1 subset}
}
\subfloat[Imaginary part of the solution obtained by using the simple randomized preconditioner.]{
\includegraphics[width=0.48\textwidth]{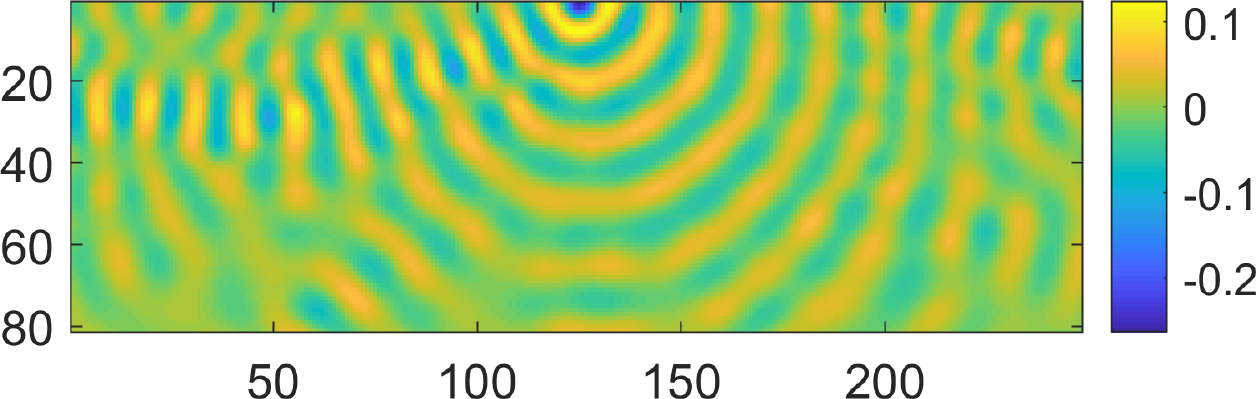}
\label{error diff 1 subset}
}\\
\subfloat[Real part of the difference]{
\includegraphics[width=0.48\textwidth]{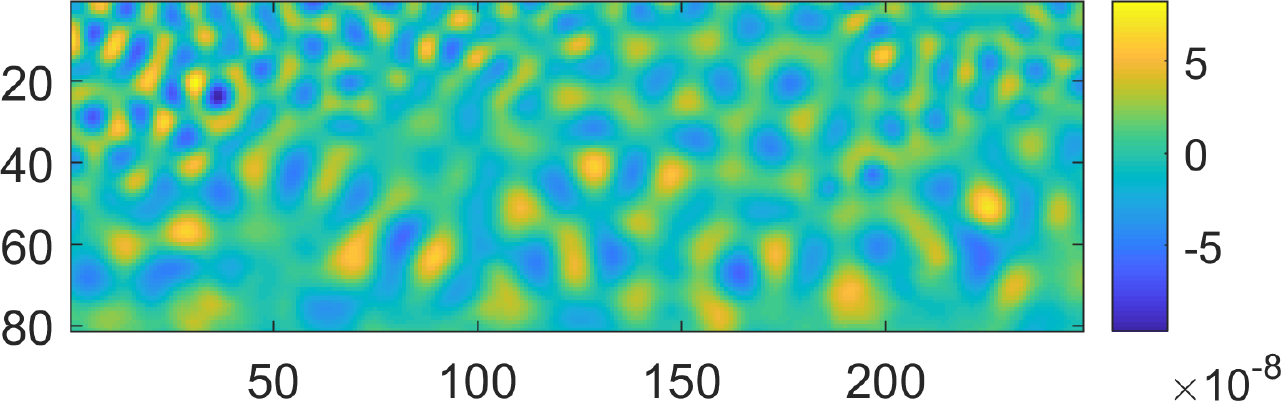}
\label{estimated difference 4}
}
\subfloat[Imaginary part of the difference]{
\includegraphics[width=0.48\textwidth]{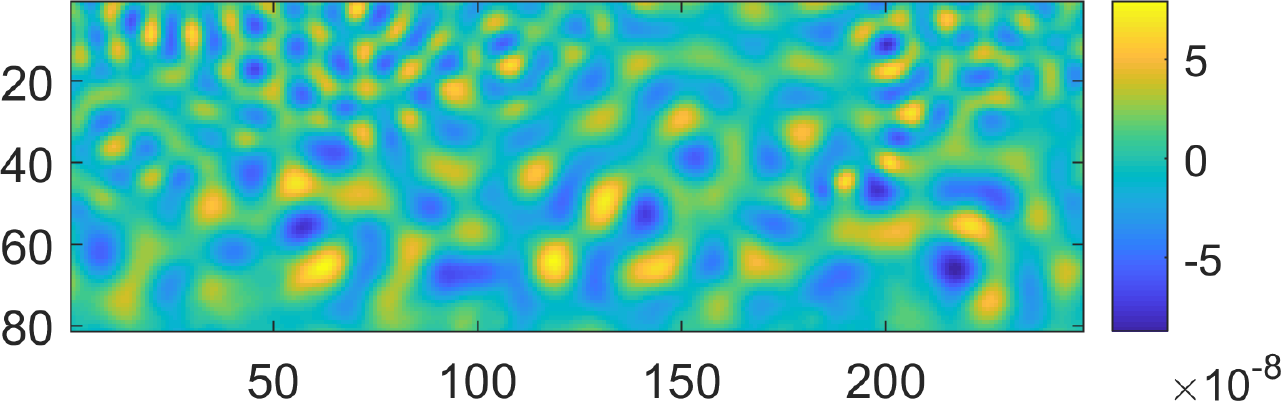}
\label{error diff 4}
}
\caption{Results for the Marmousi2 model  for 10 Hz. Comparison of the solution by \eqref{full} and iterative solution with the simple randomized preconditioner. }
\label{difference}
\end{figure*}

We used Algorithm \ref{random algorithm} to construct a preconditioner by decomposing $G^0V$ as described in section~\ref{random section}. With suitable choices of the rank $r$,   
convergence of \eqref{update2} and \eqref{update3} was obtained in few iterations, and the solution agreed with the solution obtained by solving \eqref{full}. 
Fig. \ref{difference} shows a comparison of the results of the iterative solution 
with solving \eqref{full} for 10 Hz. The results for the other frequencies were of similar quality.

\begin{figure*}
\centering
\subfloat[]{
\includegraphics[width=0.48\textwidth]{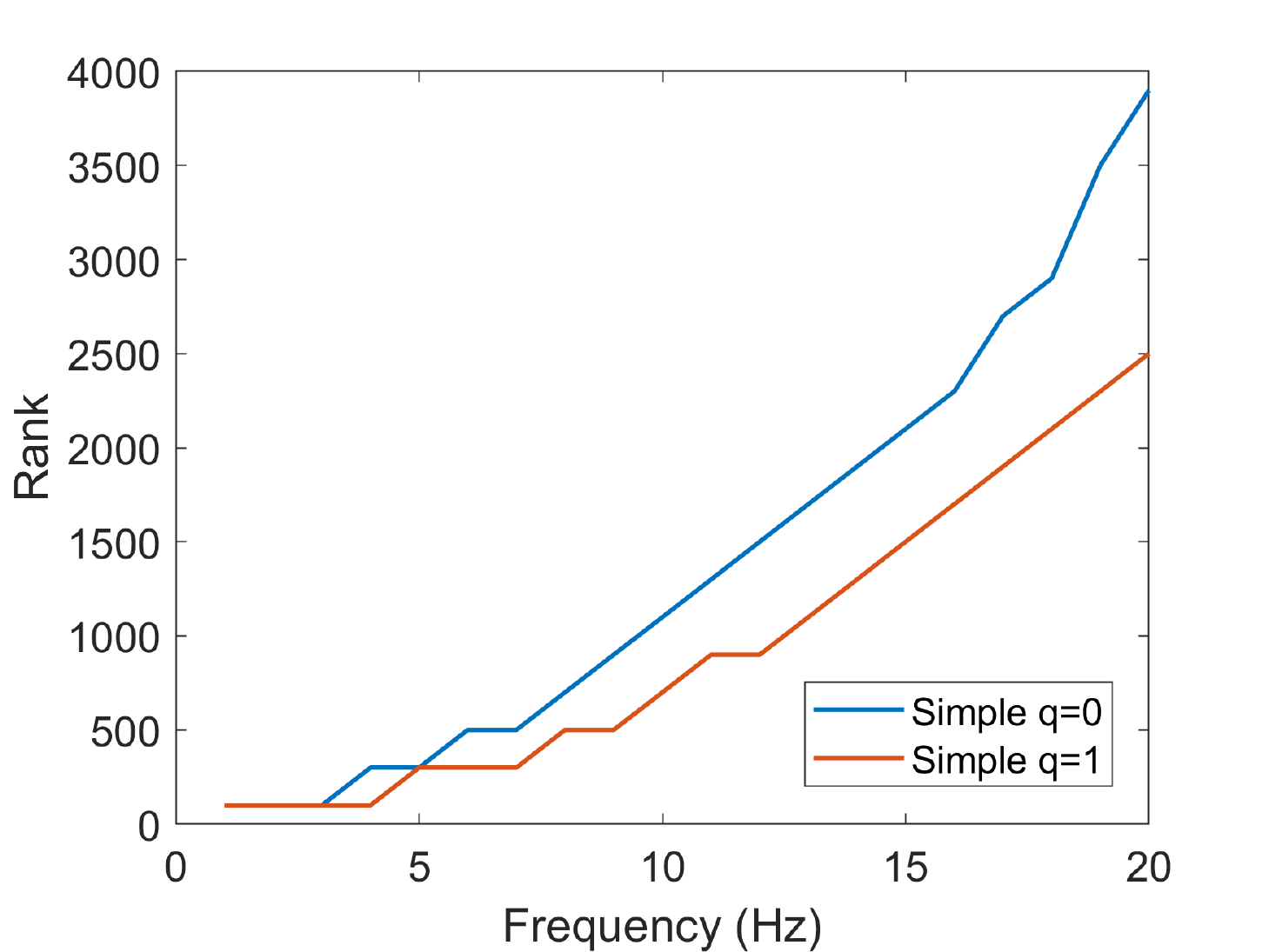}
\label{rk small}
}
\subfloat[]{
\includegraphics[width=0.48\textwidth]{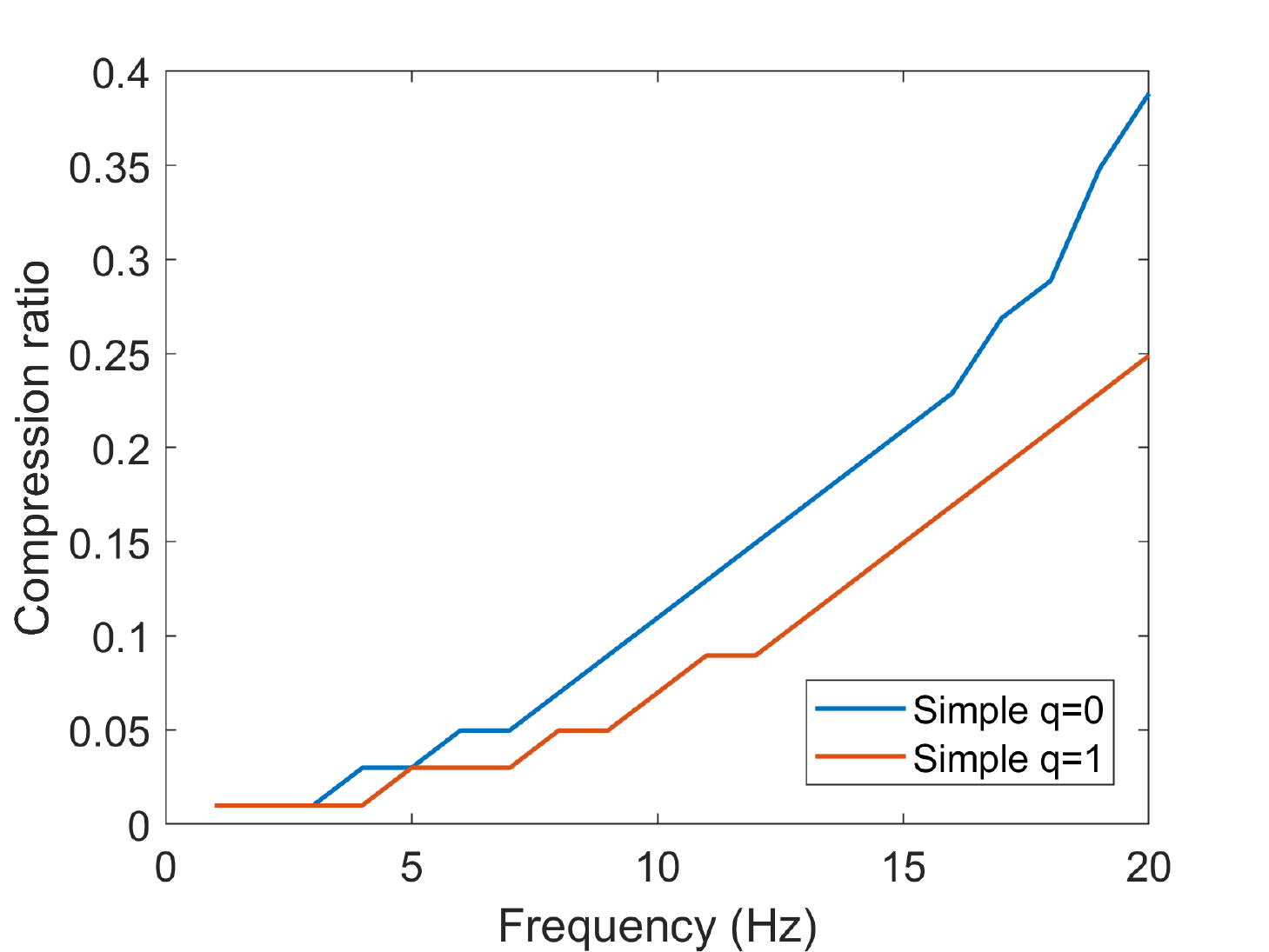}
\label{comp small}
}
\caption{(a) The rank $r$ of the approximation from Algorithm \ref{random algorithm} versus the frequency for the experiment with the simple preconditioner and the subset of the Marmousi2 
model. (b) Compression ratio.
}
\label{rc small}
\end{figure*}

\begin{figure*}
\centering
\includegraphics[width=0.48\textwidth]{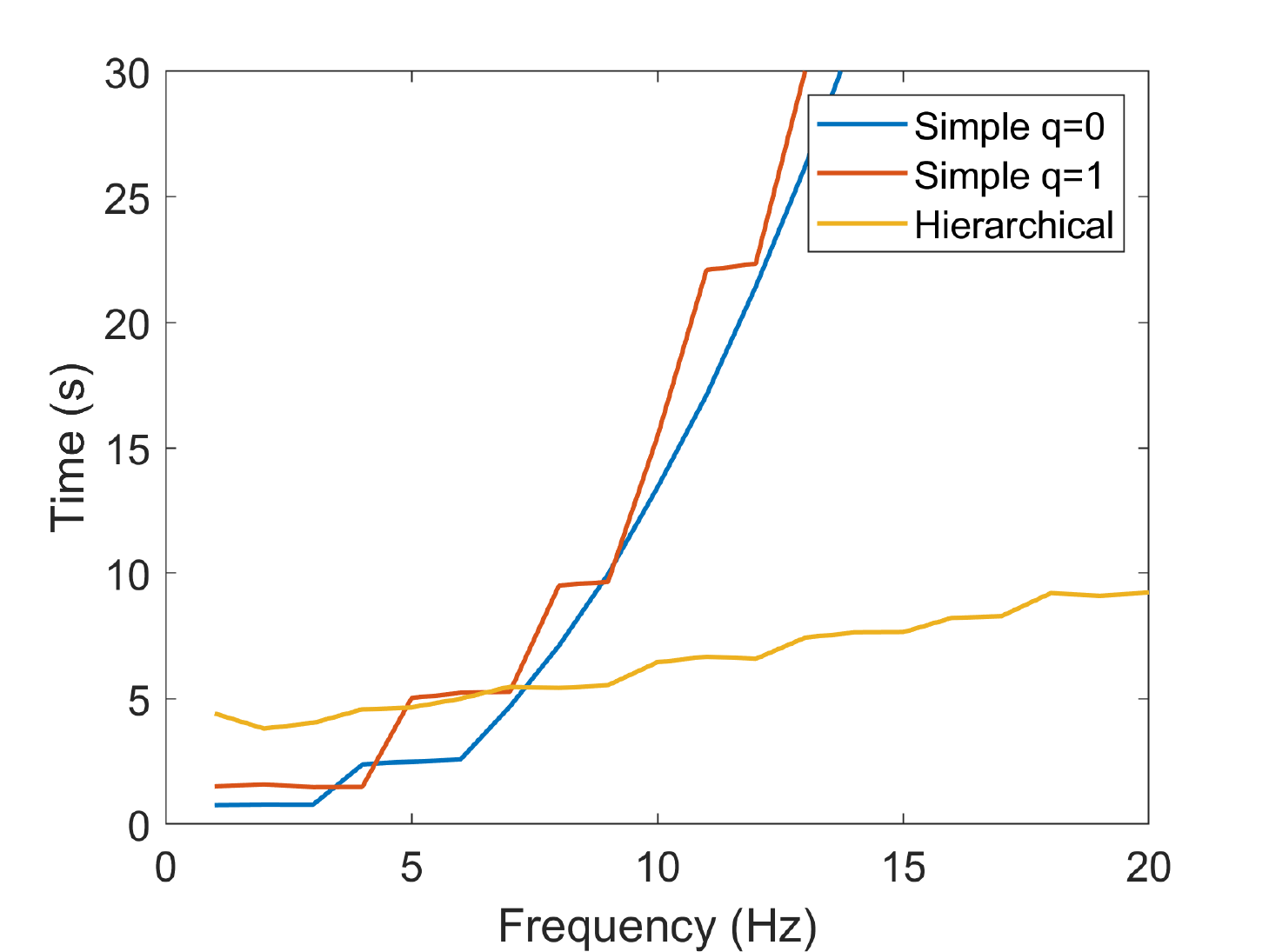}
\caption{Time used for forward modelling on the Marmousi2 model with one source for different preconditioners. The time is a total time used to construct the preconditioners plus the time for the iterations of the scattering series until convergence.The  red and blue lines show the simple randomized preconditioner with $q=0$ and $q=1$, respectively, and the yellow shows the hierarchical preconditioner. 
The simple preconditioner is fastest up to 7 Hz.  The time for the higher frequencies with the simple method is outside the range of the figure in order to
show the other results more clearly.}
\label{time}
\end{figure*}

Fig. \ref{rk small}  shows the value of the rank $r$ depending on the frequency.  
As can be seen from the figure, the required rank $r$ increases with frequency. We started with an initial value of 100 for 1 Hz and increased the rank  by 200 for the next frequency whenever more than 10 iterations of formula \eqref{update2} and \eqref{update3} were used.
We compared using $q=0,1,2$ in Algorithm \ref{random algorithm}. Only $q=0,1$ is shown in the figure, as $q=2$ gave similar results as $q=1$. It can be seen from the figure that using $q=1$ increases the accuracy, and makes it possible to use a smaller rank, but the time spent were slightly larger, see Fig. \ref{time}. Here  we used only one source, but for many sources it might be faster to use $q=1$, since larger $r$ increases the time of each iteration in \eqref{update3} a little. Fig. \ref{comp small} shows how much the preconditioner is compressed compared to the full matrix $G^0V$. For the simple preconditioner the compression ratio is $2rN/N^2=2r/N$.

For the lower frequencies the method is efficient and the necessary rank of the preconditioner is much lower than the original size of the matrix of around 20000. For the higher frequencies the performance is not as good, and we will see that the hierarchical method is better.

\subsubsection{Hierarchical preconditioner}
We use the method described in section \ref{hierarchical section} to construct a hierarchical matrix that approximates $I-G^0V$ and perform an approximate inversion. The hierarchical matrix obtained after inversion is used as H in \eqref{update}. We used 5 levels of the tree-structure as shown in Fig.~\ref{matrix}. 
All off-diagonal blocks were approximated, and the remaining squares on the diagonal were kept as full matrices.

\begin{figure*}
\centering
\subfloat[]{
\includegraphics[width=0.48\textwidth]{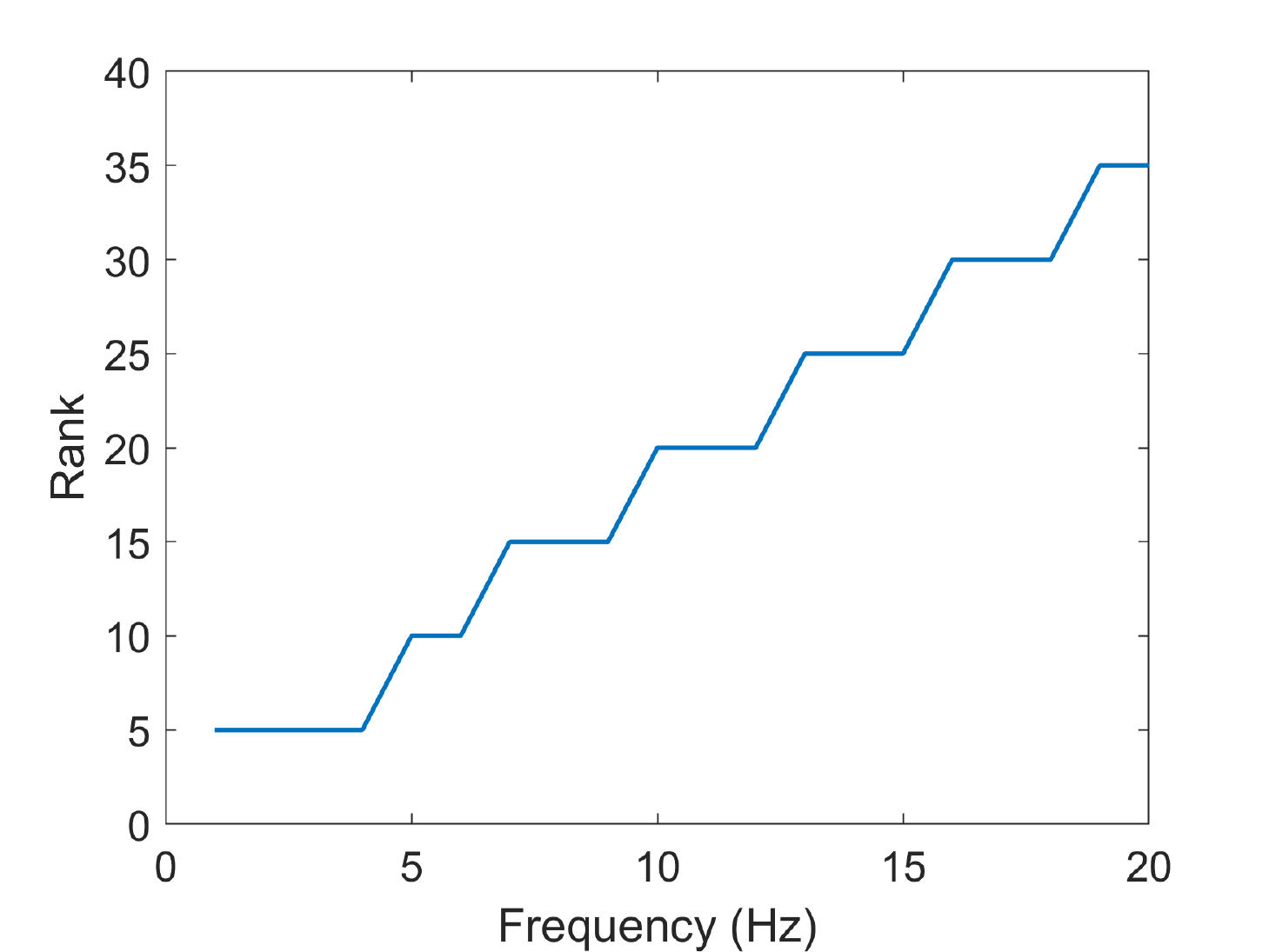}
\label{rk small hi}
}
\subfloat[]{
\includegraphics[width=0.48\textwidth]{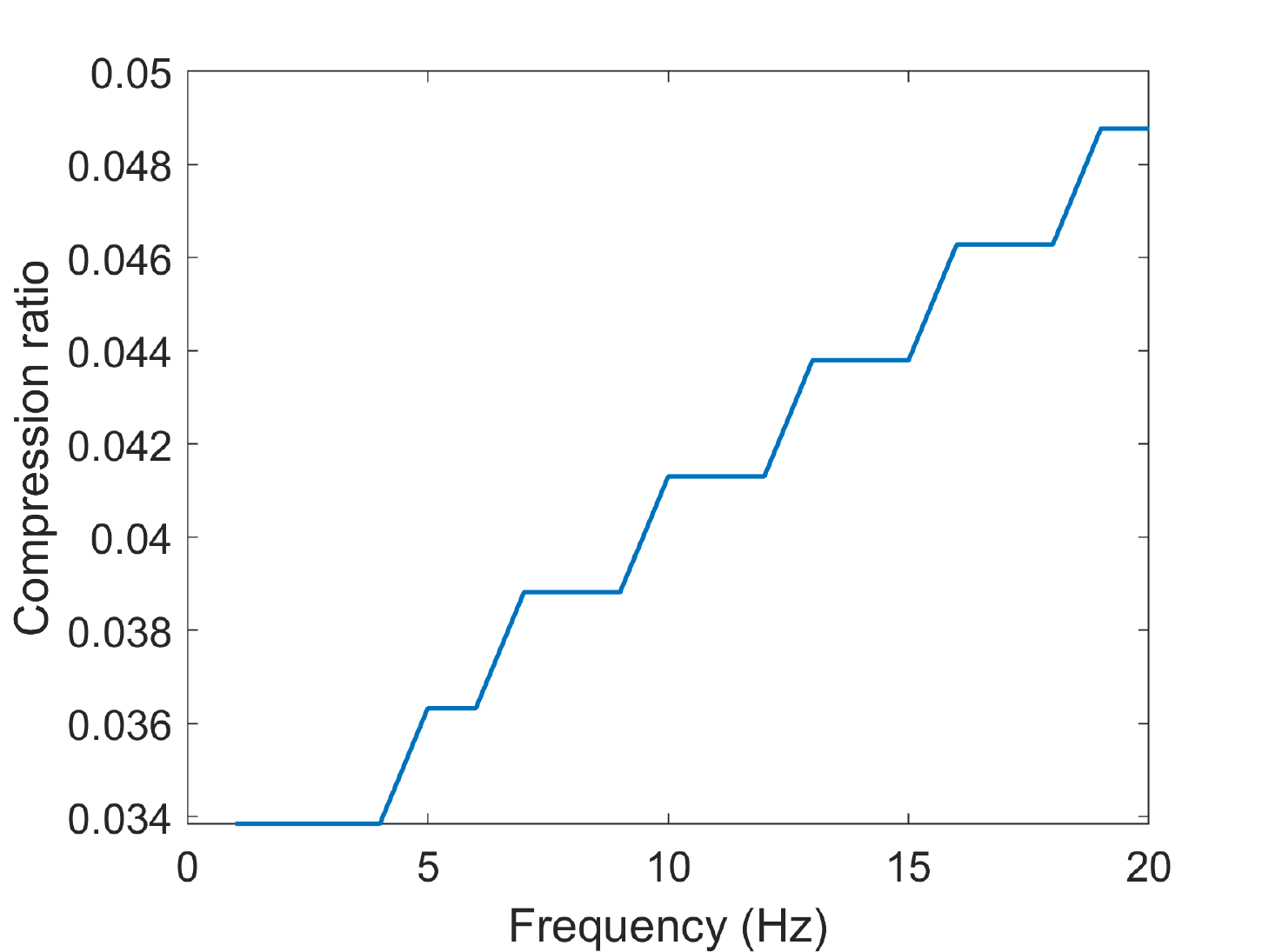}
\label{comp small hi}
}
\caption{(a) The figure shows the rank $r$ of the subblocks of the hierarchical matrix approximation  versus the frequency in the experiment with the subset of the Marmousi2 model in Fig.~\ref{small model}. (b) Compression ratio.
}
\label{rc small hi}
\end{figure*}

Fig.  \ref{rk small hi} shows the value of the rank $r$ of the off-diagonal blocks depending on the frequency in the experiment with the Marmousi2 model. As can be seen from the figure, the rank $r$ used in the subblocks of the hierarchical matrix is much lower than the rank of the simple preconditioner. But the ranks are not directly comparable since the simple method only uses one low rank approximation for the full matrix $G^0V$, and the hierarchical method has many smaller approximations. The compression ratio is shown in Fig. \ref{comp small hi}. The hierarchical preconditioner clearly gives better compression than the simple low rank preconditioner for most of the frequencies.

\begin{figure*}
\centering
\includegraphics[width=0.48\textwidth]{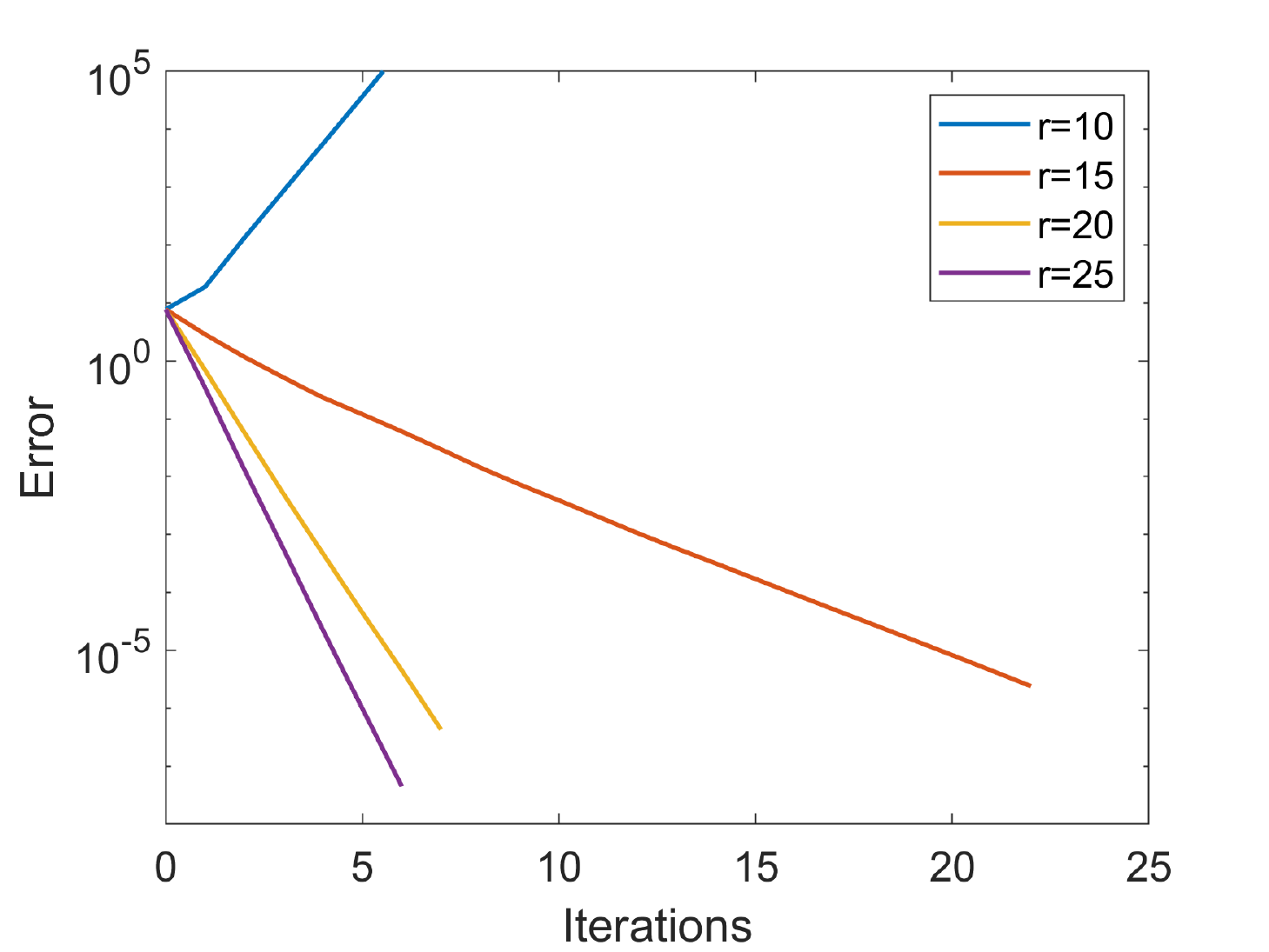}
\caption{
The figure shows how  increasing the rank of the subblocks of the hierarchical preconditioner makes the series convergent, and that higher rank gives faster convergence. The x-axis shows the number of iterations with \eqref{update}. This test is done for  the subset of the Marmousi2 model for 10 Hz. }
\label{iter}
\end{figure*}

When comparing the computational time of the two methods, we noticed that the simple method was fastest for the lower frequencies, up to 7 Hz, see Fig. \ref{time}. 
 For higher frequencies the hierarchical matrix method was clearly faster. For both methods most of the computational time was spent obtaining the preconditioner,  
and after that only a few iterations of \eqref{update} were needed for convergence (usually around 5-15). This means that the methods are well suited for applications with multiple sources, since extra sources do not require much extra computational time. The same preconditioner could be used for all sources.
Fig. \ref{iter} illustrates how the rank of the preconditioner affects the convergence of the series, and that it could be beneficial to increase the rank if there are many sources.


\subsection{Application to a larger model}
\begin{figure*}
\centering
\includegraphics[width=0.6\textwidth]{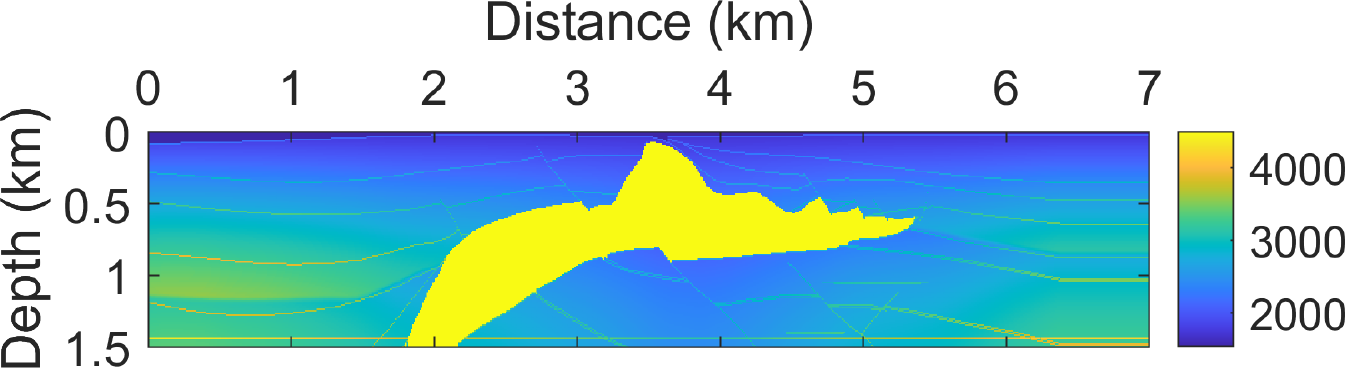}
\caption{The 2D SEG/EAGE salt model.}
\label{big model}
\end{figure*}

As a second test model we use the 2D SEG/EAGE salt model, see Fig. \ref{big model}. The number of  grid blocks is $700\times 150=105 000$ and we use a size of the grid blocks of 10 m in both directions. We assume here as well that the surroundings of the model have velocity 2000 m/s. 
We use the integer frequencies from 1 - 20 Hz. For this model the Born series is divergent for all the selected frequencies.	We test the scattering series with the two preconditioners, and then compare with using GMRES to solve \eqref{solve}. Because we compare with GMRES, we use the same stopping criteria for the scattering series and GMRES, namely $||(\psi_j-G^0V\psi_{j})-\psi_0||<10^{-6}$. 

\subsubsection{Simple preconditioner}
\begin{figure*}
\centering
\subfloat[]{
\includegraphics[width=0.48\textwidth]{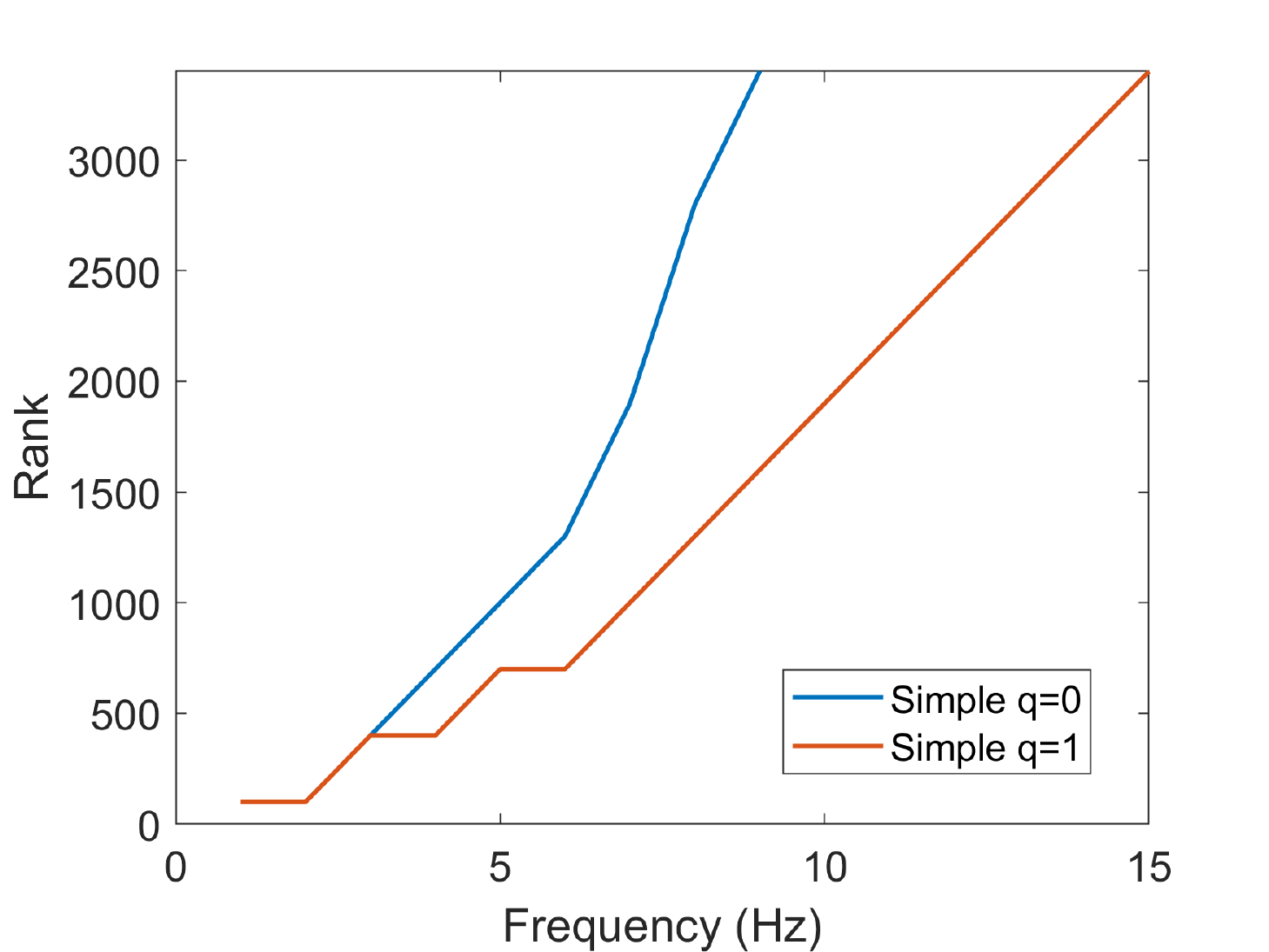}
\label{rk big}
}
\subfloat[]{
\includegraphics[width=0.48\textwidth]{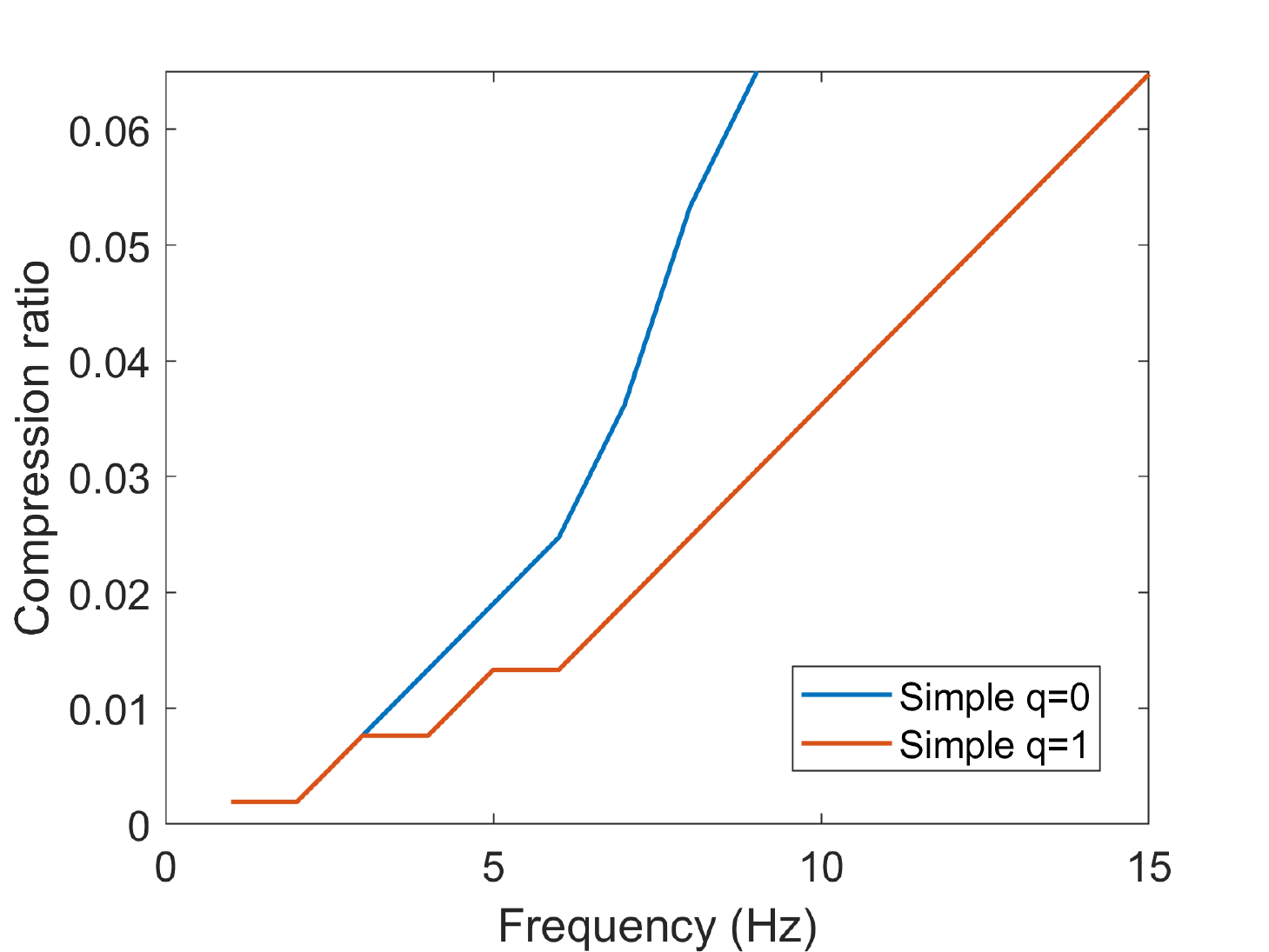}
\label{comp big}
}
\caption{(a) The rank $r$ of the simple preconditioner from Algorithm \ref{random algorithm} versus the frequency for the experiment with the salt model. The blue shows $q=0$ and the red is $q=1$. We are only showing the results for the lower frequencies, as for 
the lowest frequencies the method is efficient with r much smaller than the number of grid blocks (105000), but for the largest it is not, and the hierarchical method is better.   (b) Compression ratio.
}
\label{rc big}
\end{figure*}

\begin{figure*}
\centering
\includegraphics[width=0.48\textwidth]{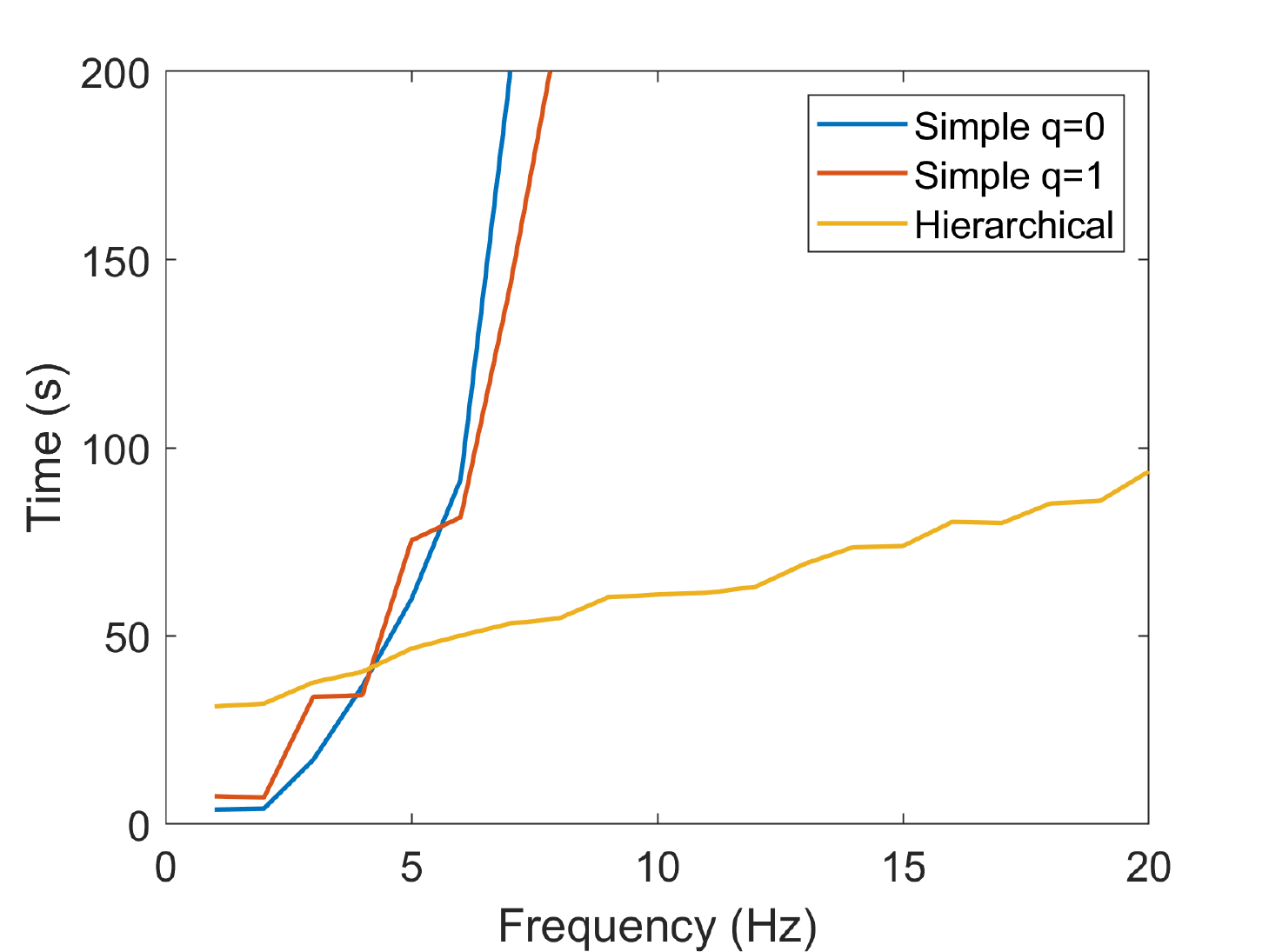}
\caption{Time used for forward modelling for the salt model with one source for different preconditioners. The time is the total time used to construct the preconditioners plus the time for the iterations of the scattering series until convergence. The blue shows the simple randomized preconditioner with $q=0$ and the red shows $q=1$, and the yellow is the hierarchical preconditioner. The majority of the time is spent on obtaining the preconditioners, so extra sources would not increase the time very much. The simple preconditioner is fastest up to 5 Hz, but for higher frequencies the hierarchical preconditioner is clearly better.}
\label{time2}
\end{figure*}

Fig. \ref{rc big}  shows the value of $r$ depending on the frequency and the compression ratio in the experiment with the salt model. 	Clearly the method is only efficient for the lower frequencies, as the rank becomes very large for the higher frequencies. The computational time for the lowest frequencies is shown in Fig. \ref{time2}. The time for the higher frequencies is outside the range of the figure in order to show the other results more clearly.

\subsubsection{Hierarchical preconditioner}
\begin{figure*}
\centering
\subfloat[]{
\includegraphics[width=0.48\textwidth]{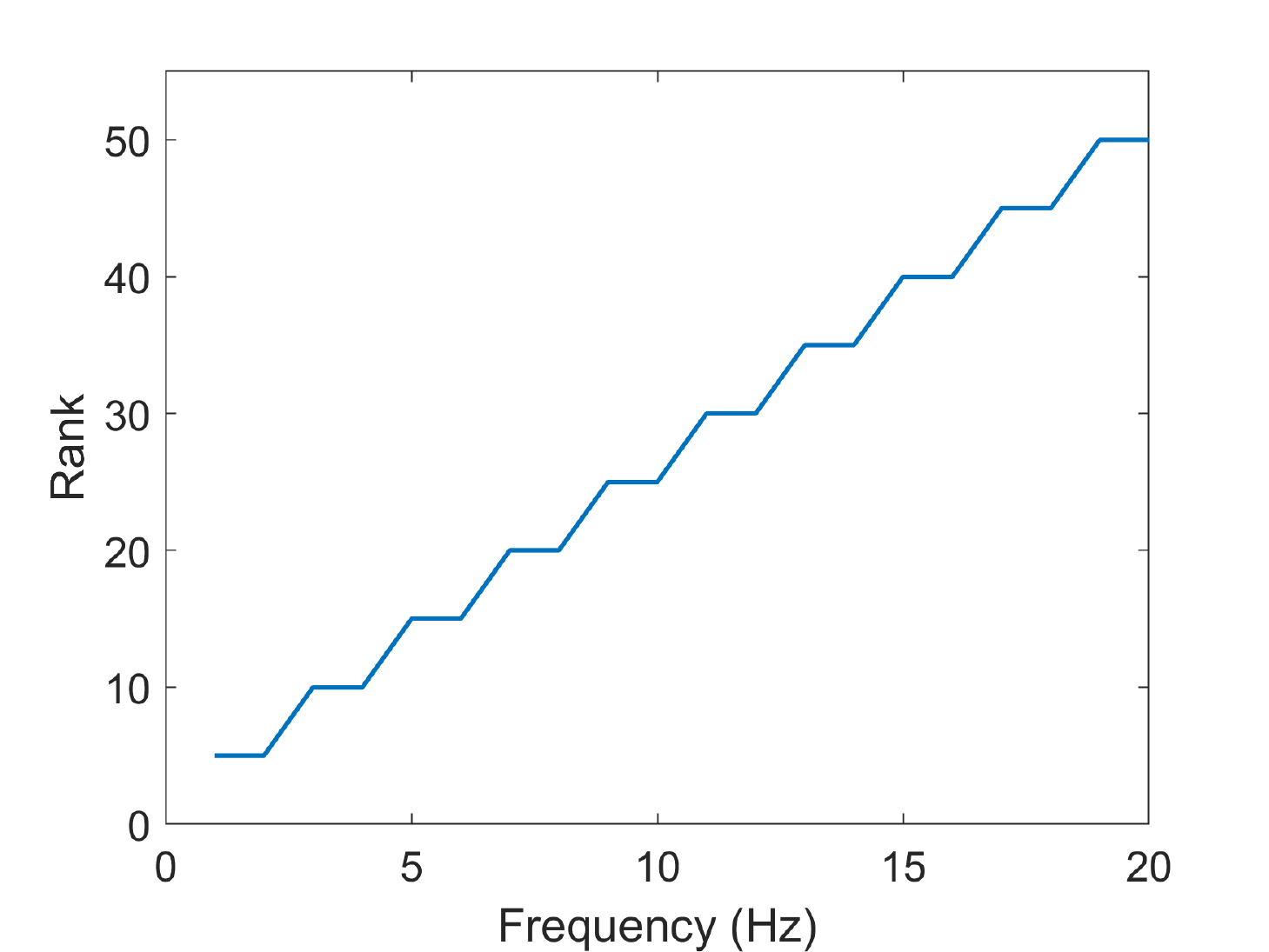}
}
\subfloat[]{
\includegraphics[width=0.48\textwidth]{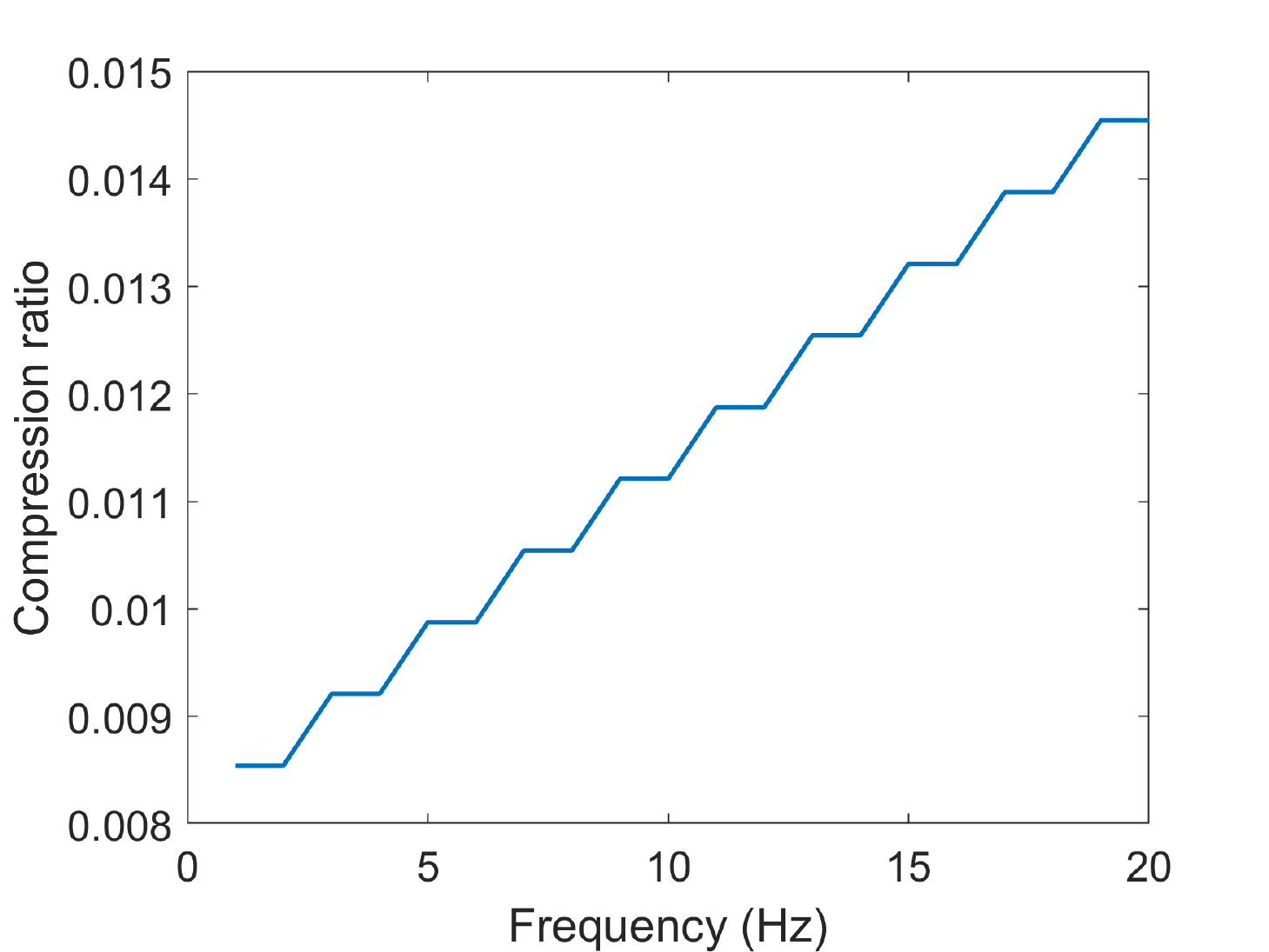}
}
\caption{(a) The rank $r$ of the subblocks of the hierarchical matrix approximation  versus the frequency in the experiment with the salt model.  (b) Compression ratio.
}
\label{rc big hi}
\end{figure*}

We used 7 levels of the tree-structure for the salt model, two more than shown in Fig.~\ref{matrix}. 
Fig.  \ref{rc big hi} shows the value of the rank $r$ and the compression ratio versus the frequency.  
We started with $r=5$ for 1~Hz and increased it with 5 for the next frequency whenever more than 10 iterations of \eqref{update} were needed for convergence. The computational time is shown in Fig. \ref{time2}. The hierarchical preconditioner is clearly better for frequencies higher than 5 Hz. 

\subsubsection{Comparison with GMRES}
\begin{figure*}
\centering
\includegraphics[width=0.48\textwidth]{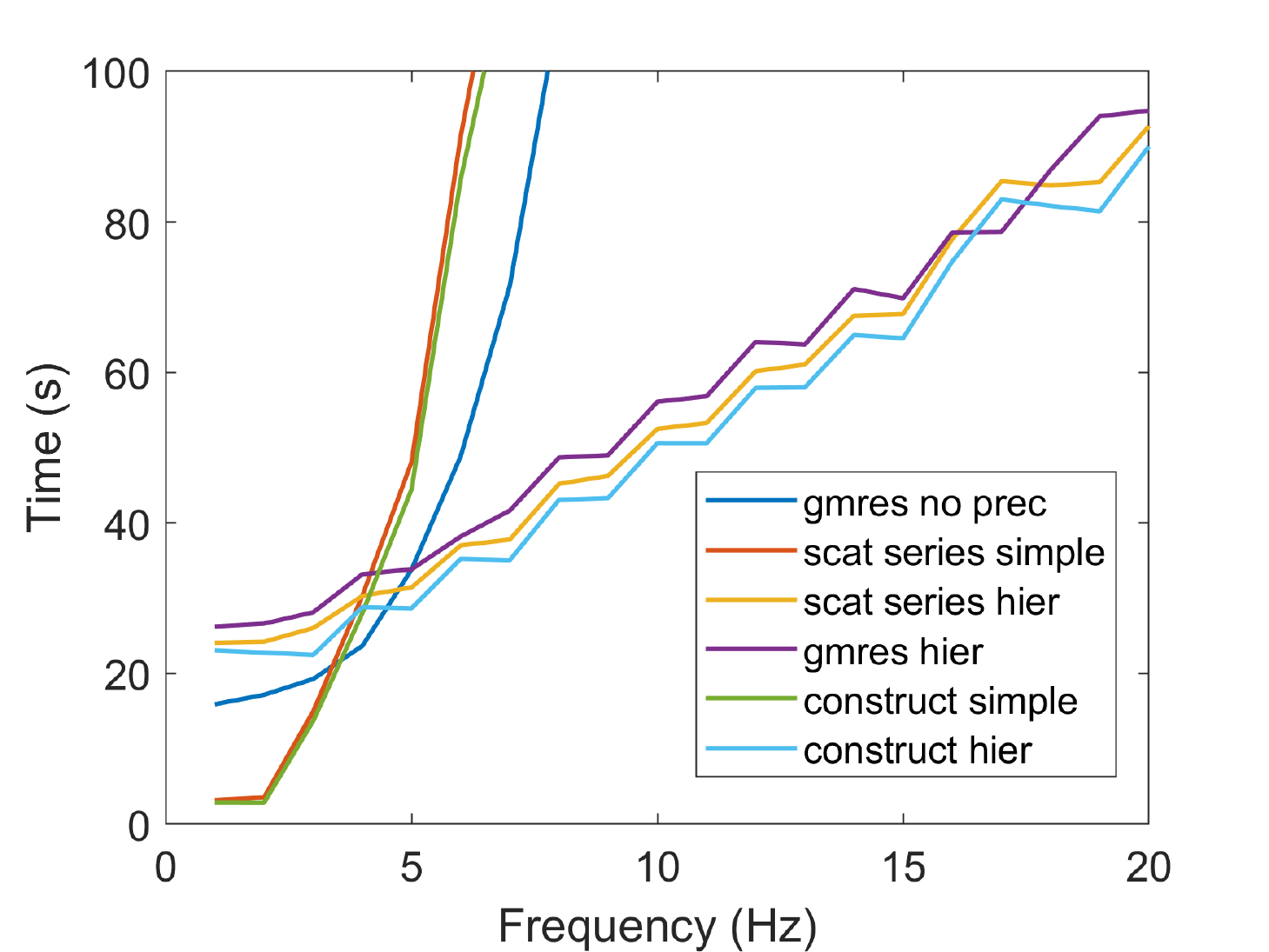}
\caption{The dark blue line shows the time used to solve \eqref{solve} for the salt model with GMRES  without a preconditioner for one source. The red and yellow lines show the time used for the scattering series with a simple preconditioner and the hierarchical preconditioner, respectively . (This was also in Fig. \ref{time2} but is repeated for comparison.) 
The purple line shows GMRES with the hierarchical preconditioner. The green and light blue lines show the time to construct the two preconditioners for the scattering series.}
\label{time3}
\end{figure*}

Although the main focus of this paper is to construct convergent scattering series, we show a comparison with the Krylov subspace method GMRES \cite[]{saad1986gmres} with and without a preconditioner to illustrate that the scattering series are efficient. We solve  
\eqref{solve} using GMRES, and FFT is used in the computation of $G^0$ times vectors. The computational time of GMRES is shown in Fig \ref{time3}. 

As can be seen from the figure, the computational time of the unpreconditioned GMRES increases quickly with frequency,  and the scattering series with the hierarchical preconditioner is clearly better for most of the frequencies. For the lowest frequencies, the simple preconditioner is the fastest. We also tested GMRES with the hierarchical  preconditioner. We used a similar procedure as for the scattering series to choose the rank of the subblocks, by increasing the rank of the  hierarchical matrix for the next frequency whenever more than 10 iterations were used. For most of the frequencies the scattering series is slightly faster than GMRES. This comparison is for one source, but the benefit of the preconditioners will be much larger  when there are several sources, as is typical in seismic applications. The time to construct the preconditioners for the scattering series is also shown in Fig.~\ref{time3}. The construction of the preconditioners takes most of the computational time, and extra sources will therefore not increase the computational time very much since the same preconditioner can be used for all sources. 


\section{Conclusion}
\label{conclusion}
We have presented  methods for solving the Lippmann-Schwinger equation in 2D in a fast and accurate way. By randomized techniques and hierarchical matrices we obtain the solution by making a scattering series  convergent. 
We presented two methods for obtaining a preconditioner for the scattering series, one where 
the Green's function times the contrast is approximated by low rank matrices, and another where we construct the approximation in a hierarchical manner. For low frequencies the first method performed well and was faster than the hierarchical method, but for the higher frequencies and in particular for the larger model, the hierarchical method performed the best. Even for low frequencies the hierarchical method was almost as good as the simple method, but as the simple method has the advantage of being very easy 
to implement,  we have described both.

Both methods are well suited for applications with multiple sources, since the majority of the computational time is spent on obtaining the preconditioner, and when it is constructed, it can be applied to several sources with little extra cost since the scattering series converges in few iterations. In this  work we focused on the  Helmholtz equation, but in principle it should be possible to extend  it to other equations that can be expressed as Lippmann-Schwinger type equations.	A more advanced structure of the hierarchical preconditioner could possibly improve the computational time further, for example by using $\mathcal{H}^2$-matrices. 
We demonstrated the methods on 2D models, but by instead using 3D FFT in the construction of the preconditioners and in the scattering series, the same can be done in 3D.
We believe the methods could also be useful for ultrasound, electromagnetic imaging and other scattering problems.  


\section*{Acknowledgements}
The authors acknowledge the Research Council of Norway and the industry partners, ConocoPhillips Skandinavia AS, Aker BP ASA, Vår Energi AS, Equinor ASA, Neptune Energy Norge AS, Lundin Norway AS, Halliburton AS, Schlumberger Norge AS, and Wintershall DEA, of The National IOR Centre of Norway for support.
The authors were also supported by the Petromaks II project 267769 (Bayesian inversion of 4D seismic waveform data for quantitative integration with production data).



\end{document}